\documentclass[12pt]{article}

    \newcommand{\na}{\nabla}
\newcommand{\ga}{\gamma}
\newcommand{\am}{a^{-1}}
\newcommand{\ampe}{a^{-1}_\perp}
\newcommand{\ampa}{a^{-1}_\parallel}
\newcommand{\gam}{\gamma^{-1}}

\newcommand{\pa}{\parallel}
\newcommand{\per}{\perp}
\newcommand{\defi}{\stackrel{df}{=}}

\title{Electrodynamics under a Possible Alternative to the Lorentz Transformation}

\author{G. D. Puccini \thanks{Present address:
Instituto de Neurociencias, Universidad Miguel Hern\'andez, Apartado
18, 03550 San Juan de Alicante, Spain.} \cr \small{Dipartimento di
    Fisica, Universit\`a di Bari, Via Amendola 173, I-70126 Bari, Italy.}}

\date{}
\begin{document}
\maketitle

\begin{abstract}
A generalization of the classical electrodynamics for systems in
  absolute motion is presented using a possible alternative to the
  Lorentz transformation.  The main hypothesis assumed in this
  work are: a) The inertial transformations relate two inertial
  frames: the privileged frame $S$ and the moving frame $S'$ with
  velocity ${\bf v}$ with respect to $S$. b) The transformation of the
  fields from $S$ to the moving frame $S'$ is given by ${\bf
  H'}=a({\bf H}- {\bf v}\times{\bf D})$ and ${\bf E'}=a({\bf E} + {\bf
  v}\times{\bf B})$ where $a$ is a matrix whose elements depend of the
  absolute velocity of the system. c) The constitutive relations in
  the moving frame $S'$ are given by ${\bf D'}= \epsilon {\bf E'}$,
  ${\bf B'}= \mu {\bf H'}$ and ${\bf J'}=\eta {\bf E'}$. It is found that Maxwell's
   equations,   which are transformed to the moving frame, take a new form depending
  on the absolute velocity of the system. Moreover, differing from
  classical electrodynamics, it is proved that the electrodynamics
  proposed explains satisfactorily the Wilson effect.
\end{abstract}


\section{Introduction}
The extension of Maxwell's electrodynamics from systems at rest to
those in motion was the fundamental problem of the beginning of
the twentieth century. In those days, it seemed impossible to
detect the absolute motion of the earth (that is, its motion with
respect to the preferred reference frame called {\em ether}) by
means of electromagnetic experiments. For the purpose of
explaining the negative results of these experiments, Lorentz's
idea was simple: if a system is in uniform translational motion
and no fundamental phenomenon is modified, this means that the
electrodynamic equations are form invariant under a certain
transformation.  Thus, Lorentz was the first to discover that
Maxwell's equations admit a transformation that leaves them
covariant.

However, it was a well-known fact that the mechanical laws did not
have the covariance property under the Lorentz
transformation. Einstein gave the decisive step introducing the
so-called {\em principle of relativity} as a restriction scheme for
natural laws. Mathematically, this principle can be expressed as
``the Lorentz covariance of {\em all} the basic laws of physics under a
change of inertial reference frame''. This covariance expresses the
group property of the transformations since the velocity of every
inertial frame is considered as {\em relative}.  In this way, it was
understood why it had been impossible to determine the ``absolute''
velocity by {\em some physical experiment}.

Nowadays, physicists agree that the Lorentz transformation describes a
fundamental symmetry of all natural phenomena. However,
it would be interesting to know if there is any alternative to such
transformation. In other words, is there any transformation between
two inertial frames which both agree with the experimental evidence
(that is, kinematically equivalent to the special relativity) and
generalize the Lorentz transformation? Or, on the contrary, is the
Lorentz transformation an unavoidable consequence of nature?

In this sense, Robertson \cite{ro49} proposed to replace the
Einstein's postulates of relativity with hypotheses suggested by
certain typical optical experiments as a way of testing the
relativity theory. Without these hypotheses, the transformations
so obtained describe a family of transformations which are
determined except for three functions depending on the
``absolute'' velocity of the reference frame. However, an implicit
hypothesis underlying these transformations is the postulate of
equality of the one-way velocity of light in all directions. A
generalization of the Robertson transformations that eliminates
this last postulate was studied by Vargas \cite{va84,va86}. This
new family of transformations, therefore, depends on four
arbitrary functions of the velocity of the reference frame. Each
member of this family is fixed with a choice of these four
functions and thus it describes a possible alternative to the
Lorentz transformation.

The most interesting member of this family of transformations was
found by Tangherlini \cite{ta61} and  studied by Mansouri \& Sexl
\cite{ma77}, Chang \cite{ch79,ch79b} and Rembieli\'nski \cite{re80}
among others. Particularly, in this transformation the
four-dimensional line element is considered as an invariant
\cite{va84,ch79,ch79b,re80,re97,ch88,va89} and, therefore, the
Minkowski metric appears modified in the moving system losing its
diagonality. As a consequence, the co-variant and contra-variant components have different
properties: if the contra-variant component of the temporal coordinate
is only dilated, its co-variant component mixes space and time; while
the opposite happens with the spatial coordinate.

The expression of Maxwell's equations under this transformation
was deduced by Chang \cite{ch79, ch79b} and by Rembieli\'nski
\cite{re80}. However, a certain ambiguity for the definition of the
fields $\bf E$ and $\bf B$ appears as a consequence of using a
non-diagonal metric with a 4-line element invariant. In order to solve
this ambiguity, Rembieli\'nski \cite{re80} has supplemented
Maxwell's equations with the equation of motion for the test charge. As
we shall see, this ambiguity is absent under the so-called ``inertial
transformations'' and, therefore, no operational definition of
electromagnetic field via equation of motion for charged particles is
necessary.

The inertial transformations proposed by Selleri have the same
mathematical form that those of Tangherlini, but they admit a
different interpretation since the four-dimensional line element
is not defined and, therefore, cannot be an invariant
\cite{se95,se96}. This means that the geometrical structure of the
four-dimensional space-time is not required into the framework of the
inertial transformation and, consequently, the geometrical structure
is given by the three-dimensional Euclidean geometry. Moreover, time
is (in one sense) similar to that of classical physics because the concept
of ``absolute simultaneity'' of the Galilean physics remain valid.
\footnote{Absolute simultaneity means that two events simultaneous in $S$ (i.e.
taking place at the same $t$) are judged also simultaneous in $S'$
(and vice-versa). This property, being a consequence of the absence of space variables in
 the
transformation of time, does not imply that time is absolute. On the contrary,
 time-dilation
phenomena similar to those of the special relativity theory \cite{se96,se96b} can be
 explained
thanks to the $\beta$-dependent factor in the transformation of time (see
 eq.(\ref{trax})). }
Given these transformations, the one-way velocity of light is the same
in all directions {\em only} in the privileged frame. Therefore, in a
moving reference frame the one-way velocity of light will be different
in every direction. This situation, however, is not in conflict
with the known experimental results since the one-way velocity of
light has never been measured (see ref.\cite{se96,se96b,cr99} for a discussion
about the non-invariant one-way velocity of light). Another
characteristic, shared with the Tangherlini transformations, is that
the inertial transformations do not form a group. As a consequence,
the absolute velocity will appear in any physical law expressed in an
inertial frame (i.e. the velocity with respect to the preferred
frame). {\footnote{ It must be stressed, however, that the Lorentz
covariance and the absolute reference frame can coexist under the
Tangherlini transformations, as has been shown by Rembieli\'nski
\cite{re80,re97}. }} In spite of these facts, the inertial
transformations provide a suitable framework within which the effects
of {\em time dilatation}, {\em Fitzgerald contraction}, the
relativistic kinematics of particles, the phenomenon of aberration of
light and Doppler effect can be satisfactorily explained
\cite{se96,se96b,cr99,pu02}. Thus, inertial transformations have
become a viable alternative to the Lorentz transformation.

In this paper we generalize the classical Maxwell electrodynamics for
moving systems using the inertial transformations. We show that such
an electrodynamics may be obtained assuming that the fields transform
in a ``form similar'' to that of special relativity. Finally, we prove
that the electrodynamics proposed explains satisfactorily the Wilson
effect.

\section{Inertial transformations}

In this section we shall present the inertial transformations in their
original form as given by Selleri \cite{se95,se96} and generalize them
in a vectorial form independent of the orientation of the axes.

The inertial transformations relate two inertial frames: $S$
(privileged system) and $S'$ (moving system), and are given by
\begin{equation}
\label{trax} \cases{t'=\gamma^{-1} t \cr x'=\gamma (x - vt)\cr
  y'=y\;\;\;\;\;z'=z}
\end{equation}
where $\gamma^{-2}= 1 - \beta^2$, $\beta^2= v^2/c^2$, with $v$ the
velocity of $S'$, and $c$ the two-way velocity of light. It is
important to note that the inertial transformations are a direct
generalization of the Galilean transformation, as it can be seen doing
$\ga =1$.

The inertial transformations as given in (\ref{trax}) are restricted
to the case of parallel axis of the two coordinate systems and the
motion of $S'$ is in the direction of the $x$ axis. It is easy to
obtain a generalization which is independent of the orientation of the
spatial axes \cite{ch79}. To do so, we observe that the spatial
component will be given by the same relativistic expression, while the
temporal component will be the same as (\ref{trax}).  That is,
\begin{equation}
\label{dire} \cases{t'= \gamma^{-1} t \cr {\bf r'}= {\bf r}+ {\bf v}
  \left[\frac{\gamma -1}{v^2} {\bf r}\cdot{\bf
v} - \gamma t \right]}
\end{equation}
In order to obtain the inverse transformation, we cannot proceed
as in the relativistic case, that is, inverting the sign of the
velocity ${\bf v}$. This is due to the difference in form of the
spatial component of the direct and inverse transformations
\begin{equation}
\cases{t=\gamma t'\cr x= \gamma^{-1} x' + v \gamma t' \label{itrax}\cr
  y=y'\;\;\;\;\;z=z'}
\end{equation}
at variance with the Lorentz transformation. To obtain such a
transformation in vectorial form, we write the spatial component as
\begin{equation}
\label{teta} {\bf r}={\bf r'}+ {\bf v} {\Theta}\;\;,
\end{equation}
where, in order to determine ${\Theta}$, we must multiply scalarly the
equation (\ref{teta}) by ${\bf v}$ and replace the scalar product ${\bf
r \cdot v}$ by the expression
\begin{equation}
\label{produ}
{\bf r\cdot v}= \gam ({\bf r' \cdot v})+ v^2 \ga t'\;\;,
\end{equation}
in the resultant equation. Note that the equation (\ref{produ}) is
obtained from (\ref{dire}) multiplying scalarly ${\bf r'}$ by
${\bf v}$, and then solving ${\bf r \cdot v}$. Finally,
substituting $\Theta$ we find
\begin{equation}
\label{inve} \cases{t= \gamma t'\cr {\bf r}= {\bf r'}+ {\bf v}
  \left[\frac{\gamma^{-1} -1}{v^2} {\bf r'}\cdot{\bf
v} + \gamma t'\right] }
\end{equation}
A difference that appears in the spatial component of this inverse
transformation with respect to those obtained with the Lorentz
transformation should be noted: the exponent of $\gamma$ is $-1$,
while in the inverse Lorentz transformation is $+1$.


\section{Generalized classical electrodynamics}

As it is well-know, the general framework to deal with electromagnetic
phenomena is given by Maxwell's equations.  In a system at rest in
the privileged frame $S$, the Maxwell equations are given by:
\footnote{We will use the MKSQ system of units.}
\begin{eqnarray}
\label{maxw}
\partial_t {\bf B}= -\na \times {\bf E}\;\;, \cr
\partial_t {\bf D} + {\bf J}= \na \times {\bf H}\;\;.
\end{eqnarray}
Moreover, these equations are supplemented by
\begin{eqnarray}
\label{maxwsup}
\na \cdot {\bf B}=0\;\;,\cr \na \cdot {\bf D}= \rho\;\;,\cr
\partial_t \rho + \na \cdot {\bf J}=0\;\;,
\end{eqnarray}
with the constitutive relations (for isotropic bodies),
\begin{equation}
\label{constit}
 {\bf D}=\epsilon {\bf E}\; ,\;\;\;\;\;{\bf B}=\mu {\bf
 H}\;,\;\;\;\;\; {\bf J}=\eta {\bf E}\;\;,
\end{equation}
where $\epsilon$ is the {\em dielectric constant}, $\mu$ is the {\em
permeability}, and $\eta$ is the {\em electric conductivity} and these
will be supposed constant throughout the medium. Moreover, in empty
space we have $\eta=0$, ${\bf D}=\epsilon_0 {\bf E}$ and ${\bf
B}=\mu_0 {\bf H}$, where $\epsilon_0 \mu_0 = 1/c^2$ is the
two-way velocity of light.

In order to obtain  Maxwell's equations in the moving frame $S'$,
first we must write the transformation formulae for the spatial and
temporal derivatives. In the simplest case, given by the transformation,
(\ref{trax}), we have:
\begin{equation}
\label{gradx} \cases{\partial_t = \gamma^{-1} \partial'_t - \gamma v
  \partial'_x \cr
\partial_x = \gamma \partial'_x \cr
\partial_y = \partial'_y\;\;\;\;\; \partial_z = \partial'_z}
\end{equation}
and for the vectorial case given by (\ref{dire}), we have:
\begin{equation}
\label{grad} \cases{\partial_t = \gamma^{-1} \partial'_t - \gamma {\bf
    v \cdot \nabla'}\cr {\bf \nabla}=
(\gamma-1) \frac{{\bf v}}{v^2} {\bf v} \cdot {\bf \nabla'} +
       {\bf \nabla'}\;\;.}
\end{equation}

We note that the Galilean case is obtained with $\ga = 1$. It will be
convenient to work with a form of this equation similar to
(\ref{gradx}), but without the distinctive role of the $x$ axis. To
obtain it, we write the gradient operator of eq. (\ref{grad}) as
\[
\nabla =\frac{ \gamma-1}{v^2}[ {\bf v} ({\bf v} \cdot \nabla') - {v^2} \nabla'] + \gamma
 \nabla'\;\;,
\]
using the decomposition ${\bf \nabla}=(\nabla_\pa, \nabla_\per)$, with
the subscripts $\pa$ and $\per$ denoting parallel and perpendicular to
the velocity ${\bf v}$. Then, we find
\begin{equation}
\label{grad2} \cases{\partial_t = \gamma^{-1} \partial'_t - \gamma
  {\bf v \cdot \tilde \nabla'}\cr {\bf \nabla}=
\gamma {\bf \tilde \nabla'} }
\end{equation}
in which we have defined $\tilde \nabla' \defi ({\nabla'_\pa},{
\gamma^{-1} \nabla'_\per})$ and replaced ${\bf v \cdot \nabla'}$ with
${\bf v \cdot \tilde \nabla'}$ in the first equation.

We can see that for any vector ${\bf A}$, the temporal derivative
will be given in the moving frame $S'$ by
\[
\partial_t {\bf A}= \gam \partial'_t {\bf A} - \ga ({\bf v \cdot
  \tilde \na'}) {\bf A}\;\;.
\]

Applying vectorial algebra, it becomes
\begin{equation}
\label{veca}
\partial_t {\bf A}= \gam
\partial'_t {\bf A} - \ga \tilde \na' \times ( {\bf A} \times {\bf v})
- \ga {\bf v}
\tilde \na'\cdot{\bf A}\;\;.
\end{equation}

Now it is easy to obtain the Maxwell equations in terms of the
coordinates of the moving frame $S'$. Substituting both the expression
that is equivalent to (\ref{veca}) and the transformation of the
gradient (\ref{grad2}) in the equations (\ref{maxw}), we obtain
\begin{eqnarray}
\label{maxwt} \gam \partial'_t {\bf B} - \ga \tilde \na' \times ({\bf
  B} \times {\bf v}) - \ga {\bf v} \tilde \na'
\cdot{\bf B} = - \ga \tilde \na' \times {\bf E}\;\;,\cr \gam
\partial'_t {\bf D}  - \ga \tilde \na' \times ({\bf
D} \times {\bf v}) - \ga {\bf v} \tilde \na' \cdot {\bf D} + {\bf J} =
\ga \tilde \na' \times {\bf H}\;\;.
\end{eqnarray}

The supplementary equations transform if we apply (\ref{grad2}) directly:
\begin{eqnarray}
\label{supt} \ga \tilde \na' \cdot {\bf B}=0\;\;, \cr \ga \tilde \na'
\cdot {\bf D}=\rho \;\;, \cr \gam
\partial'_t \rho - \ga ({\bf v} \cdot \tilde \na') \rho  + \ga \tilde
\na' \cdot {\bf J}=0\;\;.
\end{eqnarray}

By considering the first two equations (\ref{supt}), we may simplify
the expressions (\ref{maxwt}) to obtain
\begin{eqnarray}
\label{maxwtt} \gam \partial'_t {\bf B} = - \ga \tilde \na' \times
({\bf E} + {\bf v} \times {\bf B})\;\;,\cr \gam
\partial'_t {\bf D} + {\bf J} - \rho {\bf v} = \ga \tilde \na' \times
({\bf H} - {\bf v} \times {\bf D})\;\;.
\end{eqnarray}

It is important to note that we have assumed {\em no transformation
form for the fields} yet. If we assume that the fields are the same in
both frames $S$ and $S'$, that is, if the following relations are
valid,
\begin{eqnarray*}
{\bf H}({\bf r},t)= {\bf H}'({\bf r}',t')\;\;,\;\;\;\; {\bf B}({\bf
  r},t)= {\bf B}'({\bf r}',t')\;\;,\cr \cr {\bf
E}({\bf r},t)= {\bf E}'({\bf r}',t')\;\;,\;\;\;\; {\bf D}({\bf r},t)=
  {\bf D}'({\bf r}',t')\;\;,
\end{eqnarray*}
the equations (\ref{maxwtt}) become, for low velocities ({\em i.e.},
for $\ga=1$), an electrodynamics ``invariant'' under Galilean
transformations. That is, those equations express the electromagnetism
in the privileged frame as described from a moving frame with velocity
${\bf v}$. However, it is well-known that such an electrodynamics does
not agree with the experimental results. Therefore, in order to obtain
an electrodynamics in agreement with the experiment, we shall assume
that the fields do not remain invariant in both reference frames.
That is, we assume that the expressions into the rotor on the right
hand side of the equations (\ref{maxwtt}) {\em represent (except for a
matrix-coefficient) the fields in the moving frame}. In other words,
we shall postulate the following transformation laws:
%
\begin{eqnarray}
\label{hipot} {\bf H'}= a ({\bf H} - {\bf v} \times {\bf D}) \;\;,\cr \cr {\bf E'}= a
 ({\bf E} + {\bf v} \times
{\bf B}) \;\;,
\end{eqnarray}
where the matrix $a$ has the form
\[
a= \pmatrix{a_\pa(v) & 0 \cr 0 & a_\per(v)}
\]
with $a_\pa$ and $a_\per$ non nulls for any $v$, and to be
determined. The simple form assumed for the matrix $a$ is due to
the following: (a) It has to be of 2x2-dimension due to the
decomposition of the fields in their parallel and perpendicular
components. (b) It could have their four elements not equals and
depending of the value of the absolute velocity of the system. (c)
However, the non-diagonal elements are zeros due to the invariance
of the systems under rotations, because if the parallel and
perpendicular components are mixed (as it happens when the
non-diagonal elements are non nulls), this invariance would
disappear. Thus, it remains only two elements not necessarily
equals to be determined. Moreover, it should be noted that we have
assumed the same matrix for both ${\bf H'}$ and ${\bf E'}$.

Now we can obtain the transformation law for the fields ${\bf B'}$ and
${\bf D'}$. To do  this, we use the constitutive relations in vacuum, valid
not only in the privileged frame $S$ ({\em i.e.},${\bf D}= \epsilon_0
{\bf E}$ and ${\bf B}=\mu_0 {\bf H}$) but also in the moving frame
$S'$ ({\em i.e.}, ${\bf D'}= \epsilon_0 {\bf E'}$ and ${\bf B'}=\mu_0
{\bf H'}$). Multiplication by $\mu_0$ and $\epsilon_0$ in the first
and second eq. (\ref{hipot}) and the use of $\mu_0 \epsilon_0 = 1/c^2$
yield
\begin{eqnarray}
\label{invtcam2} {\bf B'} = a [{\bf B} - \frac{1}{c^2} ({\bf v} \times {\bf E})] \;\;,\cr
 \cr {\bf D'} = a [{\bf
D} + \frac{1}{c^2} ({\bf v} \times {\bf H})] \;\;.
\end{eqnarray}

The inverse transformation can be obtained with the expressions of
$\bf E$ and $\bf H$ of the equations (\ref{hipot}):
\begin{eqnarray*}
{\bf H}= \am {\bf H}' + \epsilon_0 {\bf v} \times (\am {\bf E}' - {\bf
  v} \times {\bf B}) \;\;,\cr \cr {\bf E}=
\am {\bf E}' - \mu_0 {\bf v} \times (\am {\bf H}' + {\bf v} \times
  {\bf D}) \;\;,
\end{eqnarray*}
where $a^{-1}$ is the inverse matrix of $a$ given by
\[
a^{-1}= \pmatrix{a^{-1}_\pa(v) & 0 \cr 0 & a^{-1}_\per(v)}\;\;.
\]

From the above equations, we can obtain the components which are
parallel and perpendicular to the velocity ${\bf v}$:
\begin{eqnarray}
\label{tcam1} {\bf H}_\pa = \ampa {({\bf H}' + {\bf v} \times {\bf
    D}')}_\pa \;\;,\;\;\;\; {\bf H}_\per = \ga^2
\ampe {({\bf H}' + {\bf v} \times {\bf D}')}_\per \;\;,\cr \cr
{\bf E}_\pa = \ampa {({\bf E}' - {\bf v} \times {\bf B}')}_\pa
\;\;,\;\;\;\; {\bf E}_\per = \ga^2 \ampe {({\bf E}'
- {\bf v} \times {\bf B}')}_\per \;\;,
\end{eqnarray}
where clearly ${({\bf v} \times {\bf D}')}_\pa = 0$ and ${({\bf v}
\times {\bf B}')}_\pa=0$. Using the constitutive relations in vacumm,
we have
\begin{eqnarray}
\label{tcam2} {\bf B}_\pa= \ampa {({\bf B}' + \frac{1}{c^2} {\bf v}
  \times {\bf E}')}_\pa \;\;,\;\;\;\; {\bf
B}_\per= \ga^2 \ampe {({\bf B}' + \frac{1}{c^2} {\bf v} \times {\bf
    E}')}_\per\;\;,\cr \cr
{\bf D}_\pa= \ampa {({\bf D}' - \frac{1}{c^2} {\bf v} \times {\bf
    H}')}_\pa \;\;,\;\;\;\; {\bf D}_\per= \ga^2
\ampe {({\bf D}' - \frac{1}{c^2} {\bf v} \times {\bf H}')}_\per \;\;.
\end{eqnarray}

Now we can obtain the Maxwell equations valid in the moving frame
$S'$. We calculate the second member of (\ref{maxwtt}) by
substituting the equations (\ref{hipot}):
\begin{eqnarray}
\label{segum} - \ga \tilde \na' \times ({\bf E} + {\bf v} \times {\bf
  B}) &=& -\ga a^{-1} \tilde \na' \times {\bf
E'}  \cr &=& (- a^{-1}_\per \na'_\per \times {\bf E'}_\per , -
a^{-1}_\per \ga \na'_\pa \times {\bf E'}_\per -
a^{-1}_\pa \na'_\per \times {\bf E'}_\pa )\;\;, \cr \cr \ga \tilde
\na' \times ({\bf H} - {\bf v} \times {\bf
D})&=& \ga a^{-1} \tilde \na' \times {\bf H'} \cr &=& (a^{-1}_\per
\na'_\per \times {\bf H'}_\per , a^{-1}_\per
\ga \na'_\pa \times {\bf H'}_\per + a^{-1}_\pa \na'_\per \times {\bf
  H'}_\pa )\;\;.
\end{eqnarray}
where, in the final expressions, the first component of the vector is
the parallel component.

Substituting now the expressions (\ref{tcam2}) and (\ref{segum}) in
the equations (\ref{maxwtt}), we obtain
\begin{eqnarray*}
\cases{\gam \ampa \partial'_t {\bf B}_\pa' = - a^{-1}_\per \na'_\per
  \times {\bf E'}_\per \cr \cr
\ga \ampe [\partial'_t {\bf B}_\per' + \frac{1}{c^2} \partial'_t ({\bf
    v} \times {\bf E'})_\per ]= - a^{-1}_\per
\ga \na'_\pa \times {\bf E'}_\per - a^{-1}_\pa \na'_\per \times {\bf E'}_\pa}
\end{eqnarray*}
\begin{eqnarray*}
\label{maxwprim} \cases{\gam \ampa \partial'_t {\bf D}_\pa' + ({\bf J}
    - \rho {\bf v})_\pa = a^{-1}_\per \na'_\per
\times {\bf H'}_\per \cr \cr
\ga \ampe [\partial'_t {\bf D}_\per' - \frac{1}{c^2} \partial'_t ({\bf
    v} \times {\bf H'})_\per ] + ({\bf J} -
\rho {\bf v})_\per = a^{-1}_\per \ga \na'_\pa \times {\bf H'}_\per +
    a^{-1}_\pa \na'_\per \times {\bf H'}_\pa}
\end{eqnarray*}

Moreover, in $S'$ the first two supplementary equations become,
%
\begin{eqnarray*}
\ampa \ga \na'_\pa \cdot {\bf B'}_\pa &+& \ampe \ga^2 \na'_\per \cdot
      {\bf B'}_\per  + \cr &+& \ampa \ga
\frac{1}{c^2} \na'_\pa \cdot ({\bf v} \times {\bf E'})_\pa + \ampe
      \ga^2 \frac{1}{c^2} \na'_\per \cdot ({\bf v}
\times {\bf E'})_\per = 0 \;\;,\cr \cr
\ampa \ga \na'_\pa \cdot {\bf D'}_\pa &+& \ampe \ga^2 \na'_\per \cdot
      {\bf D'}_\per +\cr &-& \ampa \ga
\frac{1}{c^2} \na'_\pa \cdot ({\bf v} \times {\bf H'})_\pa - \ampe
      \ga^2 \frac{1}{c^2} \na'_\per \cdot ({\bf v}
\times {\bf H'})_\per =\rho \;\;.
\end{eqnarray*}

The above equations acquire a simple form if we take
\begin{equation}
\label{relaa} \gam a^{-1}_\pa = a^{-1}_\per\;\;.
\end{equation}

Thus, we can write the final vectorial form of  Maxwell's equations
in the system $S'$:
\begin{eqnarray}
\label{fmaxw}
\partial'_t {\bf B'} + \frac{1}{c^2} \partial'_t ({\bf v} \times {\bf
  E'}) = - \na' \times {\bf E'}\;\;,\cr
\partial'_t {\bf D'} - \frac{1}{c^2} \partial'_t ({\bf v} \times {\bf
  H'})  + {\bf J}' = \na' \times {\bf H'}\;\;,
\end{eqnarray}
where $\bf J'$ has components ${\bf J'}_\pa \defi a_\per ({\bf J} -
\rho {\bf v})_\pa$ and ${\bf J'}_\per \defi a_\pa ({\bf J} - \rho {\bf
v})_\per$. The supplementary equations can be written in a unified
form by the use of (\ref{relaa}),
\begin{eqnarray}
\label{fsupl} \na' \cdot {\bf B'} + \frac{1}{c^2} \na' \cdot ({\bf v}
\times {\bf E'}) =0\;\;, \cr \cr \na' \cdot
{\bf D'} - \frac{1}{c^2} \na' \cdot ({\bf v} \times {\bf H'}) = a_\pa
\rho' \;\;, \cr \cr
\partial'_t \rho' + \ga \na'_\pa ({\bf J} - \rho {\bf v})_\pa +
\na'_\per ({\bf J} - \rho {\bf v})_\per= 0\;\;.
\end{eqnarray}
where we have defined, $\rho'\defi \gam \rho$. Comparing the last
equation (\ref{fsupl}) with the definitions of ${\bf J'}_\pa$ and
${\bf J'}_\per$, we can see that considering $a_\pa = 1$ and $a_\per =
\ga$, we can write finally the equation for the conservation of charge
in the frame $S'$:
\[
\partial'_t \rho' + \na' \cdot {\bf J'}=0\;\;.
\]

To complete our equations we need to know the constitutive relations
for a material medium. Here we shall assume
that the following constitutive relations are valid in the frame
$S'$:

\begin{equation}
\label{relcon}
{\bf D'}= \epsilon {\bf E'}\;\;\;,\;\;\;\;{\bf B'}= \mu {\bf H'}\;\;\;,\;\;\;\;{\bf J'}=\eta {\bf E'}
\end{equation}

That the constitutive equations have the same form of
eq.(\ref{constit}) with the same material constants $\epsilon$,
$\mu$ and $\eta$ in both $S$ and $S'$ is an assumption which can
be tested indirectly testing their consequences. In other words,
if this hypothesis is true must be inferred from the way this
framework describes the accepted knowledge and predicts results
amenable of empirical control.

With the expression of $\bf J'$ and the transformation of $\bf E'$
(eq.(\ref{hipot})), the Ohm law in the moving frame, ${\bf J'}=\eta {\bf E'}$,
becomes in the same relativistic form:

\begin{eqnarray}
\ga ({\bf J} - \rho {\bf v})_\pa = \eta ({\bf E}+  {\bf v} \times {\bf B})_\pa \;\;,\cr
 \cr
({\bf J} - \rho {\bf v})_\per = \eta \ga ({\bf E} +  {\bf v} \times {\bf B})_\per  \;\;.
\end{eqnarray}

Now we can determine the transformation laws for the fields ${\bf B'}$ and
${\bf D'}$ in a material medium by following the same steps used for the fields
in vacuum. That is, we multiply by $\mu$ and $\epsilon$ in the first and second
equation (\ref{hipot}), and then using ${\bf D}= \epsilon {\bf E}$ and
${\bf B}= \mu {\bf H}$,
 and the new assumption (\ref{relcon}) we obtain:

\begin{eqnarray}
\label{depmed}
{\bf B'}= a ({\bf B} - \mu \epsilon {\bf v} \times {\bf E}) \;\;,\cr \cr {\bf D'}= a
({\bf D} + \mu \epsilon {\bf v} \times {\bf H}) \;\;.
\end{eqnarray}

As a consequence, the transformation properties of the fields $\bf
B'$ and $\bf D'$ will depend on the properties of the medium,
which is reasonable in the present context of absolute motion.
{\footnote{Certainly this situation is different to the
relativistic case in which the constitutive relations
(\ref{relcon}) in $S'$ take a new form when is transformed to the
unprimed frame $S$ (see ref.\cite{so64}). This is due principally
to the different hypothesis assumed: In the relativistic theory we
begin from the expression for the transformation law of $\bf B'$
and $\bf H'$ (for example), and then we use the relation ${\bf
B'}=\mu {\bf H'}$ valid in $S'$ to obtain the constitutive
relation in $S$. Here, in contrast, we begin from the
eq.(\ref{hipot}) and then, with the constitutive relations in $S$
and $S'$, we obtain the transformations for $\bf B'$ and $\bf
D'$.}}

In summary, we have obtained Maxwell's equations, the supplementary
equations and the transformations of the fields for a frame in motion
with respect to the privileged frame. It is important to note that the
equations (\ref{fmaxw}) and (\ref{fsupl}) express electromagnetism as
described in the moving frame $S'$ with {\em absolute} velocity ${\bf
v}$. It should be note, moreover, that the final form of the equations
has been obtained thanks to the relation (\ref{relaa}) and to the
particular choice of the coefficients $a_\pa=1$ and $a_\per=\ga$.
With these, it is easy to see from the eqs. (\ref{hipot}) and
(\ref{tcam1}), that the fields transform in the same way as under the
Lorentz transformation. However, a such choice of the coefficients
could be wrong and the equations could take a different form. In the
following section, we shall prove that our electrodynamics is both
theoretically consistent and in agreement with the experiment (and
with the special relativity) {\em only} if the relation (\ref{relaa}) is valid.


\section{The Rowland effect}

From the expressions (\ref{fmaxw}) we can see a term (known as
``convenction current'') depending on the absolute velocity of the
inertial reference frame $S'$. This convection current $\rho {\bf v}$
produces a magnetic field (Rowland's effect), that was used by
Roentgen in 1888 in an attempt to measure the so-called ``wind of
ether'' (for a discussion of this effect see
ref.\cite{so64,ro68}). In Roentgen's experiment, the free charges
which occur at the surface of a dielectric material submitted to an
external electric field produce a magnetic field when the dielectric
is set into motion. A variant of the Roentgen experiment was proposed
by Wilson. In this experiment, a uniform magnetic field is applied
parallel to the plates of a condenser. When the dielectric is moved
perpendicularly to the magnetic field and parallel to the plates, the
condenser is charged.

The theory of relativity predicts a value for the surface charge
density which appears on the plates of
\[
\sigma_{rel}= - \ga^2 v H_0 (\epsilon \mu - \epsilon_0 \mu_0) \simeq -
v H_0 (\epsilon \mu - \epsilon_0
\mu_0)\;\;,
\]
in agreement with the experimental result \cite{ro68}. On the other
hand, it is not difficult to see that a Galilean electrodynamics
described by the equations (\ref{maxwtt}) ({\em i.e.} without
transformation of the fields) predicts:
\[
\sigma_{gal}= - v H_0 \epsilon \mu \;\;.
\]

In this section, we shall see that the surface charge density
predicted by our generalized electrodynamics correctly explains the
Wilson effect. In order to do that, let us consider a dielectric
material with dielectric constant $\epsilon$, permeability $\mu$ and
with a velocity $v$ (in the direction of the $x$ axes) with respect to
the two conductor plates (fixed in the frame $S$). Let us suppose that
the external magnetic field $H_0$ is parallel to the plates of the
condenser and perpendicular to the velocity. The surface charge
density $\sigma$ which appears on the conducting plates may be
calculated in the privileged frame $S$. We calculate the perpendicular components
of ${\bf E}'$ and ${\bf H}'$ outside the dielectric. As we know, these are given by
the relations (\ref{hipot}):

\begin{eqnarray*}
H_\per' = a_\per H_0\;\;, \cr \cr
E_\per' = a_\per \mu_0 v H_0\;\;, \cr \cr
D_\per'= \epsilon_0 E_\per' = a_\per \mu_0 \epsilon_0 v H_0\;\;.
\end{eqnarray*}

We can now calculate the expressions inside the dielectric material. Using the
second supplementary equation (\ref{fsupl}) in integral form (in $S'$):
\begin{eqnarray}
\label{ifsupl} \int_{V'} \rho' dV'=\ampa \int_{C.S.} [{\bf D'} - \frac{1}{c^2}
({\bf v} \times {\bf H'})] \cdot d{\bf S'}
\end{eqnarray}

we can obtain:

\begin{equation}
\label{sigm}
\sigma'= \ampa (a_\per \mu_0 \epsilon_0 v H_0 - D_\per')\;\;,
\end{equation}

and with the use of the relation $D_\per' = \epsilon E_\per'$ we obtain

\begin{equation}
\label{privile} E_\per' = \frac{a_\per \mu_0 \epsilon_0 v H_0 - a_\pa
 \sigma'}{\epsilon}\;\;.
\end{equation}

Hence we can find the corresponding relation in the privileged
system $S$. Substituting (\ref{privile}) in the expression
(\ref{tcam1}), $E_\per = \ga^2 \ampe({\bf E}'- {\bf v} \times {\bf
B}')_\per$, we have

\[
a_\per \ga^{-2} E_\per = \frac{a_\per \epsilon_0 \mu_0 v H_0 - a_\pa \sigma'}
{\epsilon} - a_\per \mu v H_0\;\;.
\]

Since the condenser plates are connected by a wire so that
$E_\per=0$ and taking into account that the surface charge
density, which is computed in $S'$, satisfies the relation
$\sigma'=\gam \sigma$, we finally obtain \footnote{Note that this
result can also be obtained from eq.(\ref{sigm}) replacing
$D_\per'$ by the expression given in eq.(\ref{depmed}).}

\begin{equation}
\label{conde2} \sigma = - \ampa a_\per \ga v H_0 (\epsilon \mu - \epsilon_0 \mu_0)\;\;.
\end{equation}

Therefore, this electrodynamics is in agreement with
the experiment (and with the special relativity) if we take $\ampa a_\per = \ga$. It
 should
be noted that these values have been obtained independently from the relation
 (\ref{relaa}).
In summary, we have found
the same relativistic values for the coefficients of the matrix $a$ as those proposed
in the previous section under
the theoretical criterion of symmetry in our equations.


\section{Conclusions}

We have obtained a generalization of the classical Maxwell's
electrodynamics for systems in (absolute) motion with respect to the
privileged frame. As a main hypothesis, we have assumed that the
fields transform in a ``similar'' way (as given in (\ref{hipot})) to
the relativistic case, i.e. mixing electric and magnetic fields, but
being proportional to some coefficients which are dependent on the
velocity of the reference frame.

From a theoretical point of view, we have found that the {\em most
convenient} choice of those coefficients is given by $a_\pa = 1$ and
$a_\per = \ga$. These values determine the transformation of the
fields in the {\em same} relativistic form. However, as it is expected
from transformations that do not satisfy the principle of relativity,
the Maxwell equations lose their form invariance by a term depending
on the absolute velocity of the reference frame.

As a way of testing our theory, we have solved the Wilson and we have
obtained the same relativistic result if we take $ \ampa a_\per = \ga$,
as was assumed under the theoretical basis of symmetry in our equations.
In conclusion, historically the Wilson experiment has been considered as one
of the crucial tests to reject the {\em ether} theories. However, it is
interesting to note that this effect may be explained into the Galilean
framework (as it should be expected from an effect produced at low velocities)
if certain transformation of the fields is admitted. Therefore, the property of
transformation of the fields (with $\ga=1$ in (\ref{hipot})) does not appear as
a relativistic effect, but as an effect of the motion with respect to the privileged
frame.

Finally, it should be stressed that the aim of this work was to
investigate if it is possible to obtain an electrodynamics with
different transformation properties as those of the relativistic
theory. A previous study in this line was realized by Chang
\cite{ch79,ch79b} and by Rembieli\'nski \cite{re80} who derived
the Maxwell equations under the Tangherlini transformations (in
which the 4-line element is an invariant). Rembieli\'nski
\cite{re80,re97}, moreover, have shown that the Tangherlini
transformations form a subclass of the $\bf v$-dependent
transformations of the Lorentz group. As an interesting
consequence, the Maxwell equations proposed by Chang \cite{ch79}
are obtained by Rembieli\'nski \cite{re80} by means of this $\bf
v$-transformation. Thus, into this framework the absolute
reference frame and the Lorentz covariant can coexist. In
contrast, the geometrical properties of the space and time are
very different in the inertial transformations because the 4-line
element and the metric tensor are not defined. This is the reason
why Maxwell's equations in the moving frame $S'$ have a different
form in our formulation compared with those obtained by Chang
\cite{ch79,ch79b} and by Rembieli\'nski \cite{re80}. For example,
the first Maxwell equation (\ref{maxw}) and the first
supplementary equation (\ref{maxwsup}) remain invariant in
ref.\cite{ch79,re80} whereas these equations change in our
approach. It could be interesting to investigate other possible
similarities and differences between these two frameworks.
Moreover, it is mandatory to exhibit the concordance (if any) of
these theories with other experiments. In any case, these
investigations will provide an improvement in the understanding of
the electromagnetic phenomena.


\section*{Acknowledgments}
I would like to thank the Physics Department of Bari University for the hospitality, and
 Prof. F. Selleri for many
stimulating discussions. I am also very grateful to Prof. H. Vucetich for his valuable
 comments. I would also like to
acknowledge the financial support of the Italian Government.



\end{document}